\documentclass[twocolumn,showpacs,preprintnumbers,amsmath,amssymb]{revtex4}
\usepackage{tabularx,graphicx}

\usepackage{color}
\usepackage{hyperref}
\hypersetup{
    colorlinks=true,
    linkcolor=blue,
    filecolor=blue,      
    urlcolor=blue,
}

\begin{document}

\newcommand{\beq}{\begin{equation}}
\newcommand{\eeq}{\end{equation}}
\newcommand{\beqn}{\begin{eqnarray}}
\newcommand{\eeqn}{\end{eqnarray}}
\newcommand{\bmath}{\begin{subequations}}
\newcommand{\emath}{\end{subequations}}
\newcommand{\bra}[1]{\langle #1|}
\newcommand{\ket}[1]{|#1\rangle}
\title{$\mathbf{h_\alpha}$: An index to quantify an individual's scientific leadership }
\author{J. E. Hirsch }
\address{Department of Physics, University of California, San Diego\\
La Jolla, CA 92093-0319}

\begin{abstract} 
The $\alpha$ person is the dominant person in a group. We define the $\alpha$-author of a paper as the author of the paper with the
highest $h$-index among all the coauthors, and an $\alpha$-paper of a scientist as a paper authored or coauthored by the scientist where
he/she is the $\alpha$-author. For most but not all papers in the literature there is only one $\alpha$-author.
We define the $h_\alpha$ index of a scientist as the number of papers in the $h$-core of the scientist (i.e. the set of papers
that contribute to the $h$-index of the scientist) where this scientist is the $\alpha$-author. We also define the
$h'_\alpha$ index of a scientist as the number of $\alpha$-papers of this scientist that have $\geq$
$h'_\alpha$ citations.   $h_\alpha$ and $h'_\alpha$ contain similar information, 
while $h'_\alpha$ is conceptually more appealing it  is   harder to obtain from existing databases, hence of less
current practical interest. 
We propose  that the $h_\alpha$ and/or $h'_\alpha$  indices, or other variants discussed in the paper, are  useful complements to the $h$-index of a scientist 
to quantify his/her  scientific achievement, that rectify  an inherent drawback of the
$h$-index, its inability to distinguish between authors with different coauthorships patterns.
A high $h$ index in conjunction with a high $h_\alpha/h$  ratio is a hallmark of  scientific leadership.
\end{abstract}
\maketitle 

\section{introduction}
The $h$-index 
has gained     acceptance as a bibliometric indicator of individual scientific achievement \cite{h1,h2,h3,h4}. 
Its positive features have been pointed out and analyzed. At the same time, many flaws and
shortcomings of the $h$-index have been identified and studied \cite{hc1,hc2,hc3,hc4}, and many other  bibliometric  indices have been proposed as 
alternatives to it \cite{hn1,hn2,hn3,hn4,hn5,hn6,hn7,hn8}. However, to date no other single bibliometric index has been identified that is clearly preferable to the $h$-index.
An alternative to replacing the $h$-index with another better index is to supplement 
the $h$-index with another bibliometric index that addresses
at least some of its deficiencies \cite{hs1,hs2,hs3,hs4,hs5}. To 
supplement the $h-$index, that was originally proposed as `An index to quantify an individual's scientific research output'\cite{hpaper}, in
this paper we aim to define an index to quantify an individual's scientific leadership. We should point out that this issue has been independently addressed and a solution proposed to in earlier work by 
X. Hu, R.  Rosseau and J. Chen \cite{hs6}.

Possibly the greatest shortcoming of the $h$ index is its inability to discriminate between authors that have very different coauthorship patterns. This question has been extensively addressed in the literature \cite{hs6,schr1,schr2,tsch,vavr,hh1,hh2,hh3,hh4,hh5,hh6,hh7}. How does one compare a scientist that usually publishes with one or two coauthors with another scientist that has 10 or more coauthors in every paper? Most will agree that for equally accomplished scientists a larger $h$-index is expected for the scientist with more coauthors, but how much larger? More importantly, different authors play different roles in coauthored papers. For example, consider two scientists with similar $h$-indices
where  the first one is usually the leader in the multiauthored papers he/she publishes with mostly junior coauthors, while  the 
second one is  mostly a junior coauthor in his/her multiauthored  papers. Most will agree that the first one is the more accomplished scientist, but  the $h$-indices will not reflect it.

These two aspects have been addressed in the literature in a number of important papers. 
Concerning the number of coauthors, 
Schreiber \cite{schr1,schr2, hh2}, Egghe \cite{hh1} and Gallam \cite{hh3} propose various 
algorithms to fractionally allocate credit in multiauthored papers that goes inversely with the number of
coauthors, defining new indices $h_m$, $fractional$ $h$ and $gh$ respectively as alternatives to the $h$ index.
In this author's view, one or more of these indices may well be superior to the $h$ index as a single indicator
{\it if it was as easily calculable as the h-index}. However, in this author's experience obtaining these indices is considerably  more time
consuming than obtaining the h-index.
Certainly they have merit as useful supplements to the h-index to discriminate between authors that publish
alone or in small collaborations versus  those publishing in larger collaborations. 

Concerning the different roles played by coauthors in a collaboration, this issue has   been addressed
 in works by Tscharntke \cite{tsch}, Liu et al \cite{hh4}, Ancheyta \cite{hh5}, Ausloos \cite{hh6},
Crispo \cite{hh7} and Vavrycuk \cite{vavr}. These authors propose various schemes to 
apportion different credit in a multiauthored paper based on the different roles played by different coauthors.
For example, first authors are given extra  credit, or last authors are given extra credit, or corresponding
authors are given extra credit, or individual contributions are considered as described by the authors
themselves. Reference \cite{vavr} summarizes these different schemes and proposes a 
`combined weighted counting scheme'. In this author's view, while these proposals may be very useful
as applied to a particular field or discipline, they cannot be used across the board because of the very
different practices in different disciplines regarding order of authors, significance of authorship
position in the author's list, etc.

The index $h_\alpha$ proposed in this paper addresses the two aspects of the coauthorsip issue
addressed in the body of work discussed above. We argue that it contributes to resolve these issues in a more
comprehensive and efficient way than any of the indices proposed earlier.  Of course it also has its own drawbacks that we will discuss.

 In earlier work we have proposed the $\hbar$-index (h-bar index) to address these issues \cite{hbar}. The $\hbar$-index only 
 counts the papers that contribute to the $h$-index of all its coauthors. Thus, it affects negatively authors that publish with a large number
 of coauthors and with more senior coauthors. This index has not gained wide acceptance, in part perhaps because it is difficult
 to obtain. Also, a shortcoming of the $\hbar$-index is that when a paper gains enough citations it will contribute to the $\hbar$-index
 of all its coauthors equally, independently of what the relative contribution of each coauthor to the paper was. More importantly,
 in many cases the $\hbar$-index may not be sufficiently different from the $h$-index of an author to justify the substantial extra
 work needed to obtain it.

We have emphasized from the outset that the $h$-index should be only one of many elements used in
evaluating scientific achievement of an individual \cite{hpaper}. Because it has become perhaps of outsized importance in 
the evaluation of scientists, we believe it is important to supplement it with a quantifiable assessment of the 
relative importance of the given scientist in the collaborative work that contributes to his/her $h$-index.
To do so, we propose the $h_\alpha$ index in this paper. Its name refers to the fact that the $\alpha$ person is
the dominant person in a group \cite{alphawiki}. The purpose of the $h_\alpha$ index is to give a measure  of those
highly cited scientific contributions of a scientist for which  the scientist is the dominant person in  the collaboration resulting
in  a multiauthored paper, who we will call the $\alpha$-author. In other words, the $h_\alpha$ index measures
scientific leadership.

Identifying the $\alpha$-author in a collaboration is not a trivial matter, and may not even be a well-defined question. Is it the
scientist that procured the funding, is it the most senior scientist, is it the one that provided the key idea that got the 
project started, or the one that did most of the work? In many cases these roles may be played by different coauthors, in other cases 
several coauthors play similarly important roles in these tasks. Nevertheless, we argue that in most situations it is possible
to identify a key person as the $\alpha-$person in a collaboration. For lack of a better criterion, {\it we $define$ the 
$\alpha$-author of a paper  to be the coauthor with highest $h-$index}. Because a high $h$-index is generally an
indication of high scientific achievement, we argue that this is a reasonable criterion. 
To determine who is the $\alpha-$author we use the $h-$indices at the present time, rather than the $h-$indices at the time the paper was published,
which are not available in existing databases. Under the assumption that $h-$indices grow at similar rates, both choices would give similar
results. 
Of course for single-author papers the author is the
$\alpha-$author.

We propose two  indices, which we call  $h_\alpha$ and
$h'_\alpha$ indices. They contain similar information. They are not proposed to replace
but rather to complement the $h$ index. As stated in the abstract, the $h'_\alpha$ index of a scientist is defined exactly the same way the $h$-index is defined except that it refers only to $\alpha$-papers of a scientist, i.e. those
papers where the scientist is  the $\alpha$-author.  It is possible that a paper
could have two or more $\alpha$-authors but that generally will not be the case.
So {\it a scientist with $h$-index $h$ has $h'_\alpha$-index $h'_\alpha$ if that scientist has written $h'_\alpha$ papers that have
$\geq$   $h'_\alpha$ citations each, and where all the coauthors of each of those  papers have $h$-indices lower or equal to $h$}.

The $h'_\alpha$ index as defined above is difficult to obtain from existing databases. For that reason we 
 define the related index $h_\alpha$ of a scientist as the number of papers in the $h$-core of the scientist
for which the scientist is the $\alpha$ author. It may be the case that $h_\alpha=h'_\alpha$, 
in general $h_\alpha \leq h'_\alpha$. The reason to define $h_\alpha$ is that it is easier to calculate from
existing databases. One simply has to go through the list of papers in the $h$-core of a scientist and eliminate
those papers for which a coauthor has higher $h$-index than the $h$-index of the author under consideration.

It is clear from the definition that $h_\alpha \leq h$, and we will argue that the ratio  $0\leq h_\alpha/h\leq 1$ gives 
useful information.
Note also that the set of papers that contribute to a scientist's $h'_\alpha$ index  may be a subset of the set of papers contributing to the scientist's
$h$ index, in which case $h'_\alpha=h_\alpha$, or it may have some subset of it that
belongs to  the $h-$core and another subset that does not, in which case $h'_\alpha>h_\alpha$, and it may even be the case that the $h'_\alpha$ core and the $h-$core are disjoint sets, in which case $h_\alpha=0$, $h'_\alpha>0$. 
However, the latter situation will probably be very rare except   for very junior scientists.

The $h_\alpha$ and $h'_\alpha$ indices are certainly of no use for beginning scientists, e.g. graduate students or post-docs.
Such very young scientists will  write all or almost all their papers with their advisors and hence 
those papers  will contribute
nothing to their $h_\alpha$ or $h'_\alpha$ indices. At some later point in their career this should begin to change, 
at which point first $h'_\alpha$ and then $h_\alpha$ will start to provide useful additional bibliometric  information to what their
$h-$index provides.

\section{case study}

\begin{table}
\caption{Bibliometric data for 16 physicists at the physics department of a major research university in the United States \cite{mru} (A through P), and one
physicist at Princeton University (AA), listed in order of 
increasing $h$-index. The data for the index $h_\alpha$ introduced
in this paper, and the ratio $h_\alpha/h$, are in boldface. `pubs' is the number of papers published, `years' is years from publication of first paper to 
the present, $m$ is the ratio $h/years$. The research fields of these physicists are high energy theory (het), 
condensed matter theory (cmt), plasma theory (pt), biophysics theory (bpt) and observational astrophysics (oap).
Note the very strong variations in the $h_\alpha$ index and the ratio $h_\alpha/h$. For the explanation of the
red coloring, see text.
}

\begin{tabular}{l || c | c  |c|  c| |c |c | c|c }
 Name & $h$ &$ h_\alpha$&$r_\alpha =  h_\alpha/h$& $m$ &pubs & citations &years &field   \cr
\tableline
\tableline
A &25 &    \bf{8} &\bf{0.32} & 1.39  &59&2944 &18 & het    \cr    
 \textcolor{red}{B} &27 &    \bf {19} &\bf{0.70} & 0.52 & 83&3649 & 52 & het    \cr
\tableline
 \textcolor{red}{C}&32 &   \bf{ 8}  &\bf{0.25} &1.10 & 127 &4040 & 29 &het  \cr
\tableline
D &34&   \bf{ 5 } &\bf{0.15} & 0.97 & 93 &5377& 35&cmt  \cr
\tableline
E &34 &   \bf{ 22} &\bf{0.65} &0.97 & 133 &3967 &35&  pt  \cr
\tableline
F&36 &   \bf{ 16} &\bf{0.44} &1.89 & 104 &4702 &19 & cmt \cr
\tableline
 \textcolor{red}{G}  &36 &   \bf{ 7} &\bf{0.19} &1.09 & 146 &39,062 & 33 &het  \cr
\tableline
H  &37 &   \bf{ 18} &\bf{0.49} &1.32 & 80 &6285&28& het  \cr
\tableline
I &39 &  \bf{ 4} &\bf{0.10}&1.63 & 130 &5823  & 24& cmt\cr
\tableline
J  &39 &   \bf{16} &\bf{0.41} &1.39 & 119&6582&28 & bpt \cr
\tableline
K  &40 &   \bf{2} &\bf{0.05} &1.48 & 273  &6815 &27 & oap \cr
\tableline
L  &43 &   \bf{12} &\bf{0.28} & 1.39 & 104&5631 &31 & het \cr
\tableline
 \textcolor{red}{M}  &47 &  \bf{30} &\bf{0.64} & 1.62 & 186&9943&29 & het \cr
\tableline
N &50 &  \bf{27} &\bf{0.54}& 2.17 & 268&12,536 & 23 &cmt \cr
\tableline
\tableline
\tableline
AA &55 & \bf{ 51} &\bf{0.93}& 1.31 & 116 & 23,509 &42 & cmt \cr
\tableline
\tableline
\tableline
O &60 & \bf{ 1} &\bf{0.02}& 5.45 & 160&14,190 & 11 &oap \cr
\tableline
P &60 & \bf{ 14} &\bf{0.23}& 3.16 & 224&11,068 & 19 &oap \cr
\tableline

\end{tabular}
\end{table}

We use the Web of Science for the bibliometric data, and in particular the database ResearcherId when possible \cite{rid}. ResearcherId is a very 
useful feature because it provides name disambiguation.
Table I shows publication and citation metrics for 13 theoretical physicists at the physics department of a major research university in the United States \cite{mru}, henceforth called ``MRU''
(entries A through J and L through N), three observational astrophysicists at MRU (K, O, P), and one theoretical physicist at Princeton University (AA).
The data are arranged in order of increasing $h$-index and include all theorists in the department of physics at MRU 
with $h$ in the range $25\leq h \leq 50$. The table gives the seniority of the researcher by listing the number of years
since publication of the first paper (`years'), which is usually close to (typically 1-3 years before) the Ph.D. date.

The first thing to note from table I is that there is not a strong positive correlation of the $h-$index with seniority, or
equivalently years from  Ph.D. degree.
This is of course not surprising, since different scientists produce research at different rates, and the quality and  impact of the 
research differs widely.

 Turning to $h_\alpha$, note the large differences in $h_\alpha$ for physicists with similar $h-$indices, reflecting very different coauthorship patterns and degree of scientific leadership.
 Note  also that the ratio $r_\alpha\equiv h_\alpha/h$ in table I is not strongly  correlated with `years'. In other words, more seniority
 does not necessarily lead to higher independence and scientific leadership, contrary to what might have been expected.
 This suggests  that scientific leaders start leading early on in their career.

The ratio $m=h/years$ ($years=$ years from  first published paper to the present) was defined in ref. \cite{hpaper}, where it was pointed
out that a high value of $m$ indicates `outstanding scientists'   independent of seniority. However that statement has to be
tempered when taking into account $h_\alpha$. We can see from table I that $m$ values above 1.4 are
sometimes associated with high values of  $r_\alpha=h_\alpha/h$ (F, M, N)  and sometimes with low values (I, K, O, P). 
In the latter case, particularly because  high values of $m$ may result from many papers in  large collaborations rather than
from high individual achievement, 
what was stated in ref. \cite{hpaper} quoted above obviously  does not necessarily follow.

One of the motivations for the original introduction of the $h$-index was that the alternative of considering 
total number of citations could easily lead to misleading results. 
Indeed, just looking at the ``citations'' column in table I would lead to the conclusion that physicist G is by far
the most accomplished scientist on the list, with $39,062$ citations. In fact, this high number comes about because 
physicist G coauthored 9 review articles (``Review of Particle Physics''), each   having several  thousand 
citations and several hundred coauthors.  The total number of citations of physicist G excluding those review articles is 3882, i.e. the total
citation number  is 10 times larger than the citations to non-review articles.
Instead, these multi-authored review articles augment the $h-$ index of this author by $30 \%$ only,
properly decreasing their importance. 
With the $h_\alpha$ index, the effect of these review papers, which are not really representative of this
author's scientific accomplishments, is completely eliminated, since each of the review articles has several other authors with 
(much) higher $h-$index than this author.

More generally, table I shows very little correlation between the $h$-indices and $h_\alpha$ indices. I argue that the $h_\alpha$ index
is essential information to take into account in the evaluation and comparison of these scientists. 

For example, physicists O and  P have the highest $h=60$ and highest $m$-values  but smaller $h_\alpha$ indices (1 and 14) than
physicists B, E, F, H, J, M, N and AA, sometimes substantially so.  It would be wrong to just rely on the
$h$-index to conclude that O and P are the most accomplished of the list. The reason O's and P's $h$ indices are  so high
is because a large number of their  papers are coauthored by between 10 and 40 authors, some of them with
$h$-index substantially larger than O's and P's, indicating that O and P are not the leaders in these collaborations.
Because O and P are not the $\alpha$-authors in these papers, the papers  don't contribute to O's and P's $h_\alpha$ indices,
resulting in $h_\alpha$ indices that are 60 times and 4 times smaller than their  $h$-indices respectively. Instead, 
the higher $h_\alpha$ index and $r_\alpha$ ratio of the other physicists in the comparison group reflect the fact that
they are the leading authors in a substantially larger number of their highly cited papers,
which suggests that they are more accomplished scientists.

The situation  is the same  for physicist K. With an $h=40$, 27 years after the first paper, $m=1.48$, and 6,815 total 
citations, one might have concluded that this is an outstanding scientist. However, physicist K's $h_\alpha$ is a mere
2, and the  $\alpha$-ratio $r_\alpha=0.05$. Many of the papers of physicist K are written in collaboration with 20-40 coauthors,
and both in those as well as other papers with fewer coauthors there are  coauthors with $h-$indices 
higher than K's, often substantially so. These data suggest that K is the scientific leader in only 2 out of the 40 papers in 
physicist K's  $h-$core.

Comparing   high energy theorists B and L, 
it would be  reasonable to  conclude that physicist B, with an $h$ index of only 27 and $h_\alpha$ of 19, is more
accomplished than  physicist L, with h=43 but $h_\alpha$ only 12, contrary to what their relative $h-$indices
suggest. Indeed, B has the rank of Distinguished Professor in the Department, while L has the lower  rank of Professor.
 
Similarly, in a comparison between condensed matter theorists, one would reasonably conclude that
physicist F, with h=36 and $h_\alpha=16$, is more accomplished than I, with 
$h=39$ but $h_\alpha$ only $4$. Physicist I writes many papers in large collaborations with high-$h$ scientists and it would be
hard to believe  that physicist I is the leader in these collaborations. His/her low $h_\alpha$ index properly
reflects this fact. In contrast, F is the scientific leader in a substantial fraction of his/her papers which are coauthored
with his/her students and postdocs.

Comparing theorists D and E, assuming this is possible even though they are in different physics subfields,
we learn that they have the same number of years since their first paper (35) and the same $h-$index (34), and D has somewhat more total citations than
E (5377 versus 3967). One might have concluded from this information that D and E are similar, D somewhat more accomplished. However,
their $h_\alpha$ index differs by a factor of 4 (5 and 22), with E having the higher one. This indicates that the $h-$index of E results
in large part from independent work where he/she is the leader, and that of D from collaborative work with 
more senior scientists where D  is not likely to have  played the leading role. Both D and E work  in small collaborations 
involving at most a few coauthors.

There are 12 physicists on the list of table I that hold the rank of Professor in the department of physics at MRU,
and 4 that hold the higher rank of Distinguished Professor \cite{dist}. 
The latter ones are B, C, G, and M, colored  \textcolor{red}{red} in the table. Could one have inferred this from the data given in table I?
The answer is clearly no. To begin with, the Distinguished Professors  are certainly not the ones with highest $h-$indices.
Taking into account $h_\alpha$,  the data in table I would suggests that if C and G are  at the highest rank,
E and F, that have comparable $h$-indices to C and G but substantially higher $h_\alpha$ indices, should certainly 
be at the highest rank, but they are not.  E has also more seniority (35)  than C and G (29 and 33). 
Similarly, while it seems clearly justified that M is a Distinguished Professor given his/her high
$h$ and $h_\alpha$ while  L, O and P, with comparable $h-$indices but much lower $h_\alpha$-indices are not, it is surprising that $N$, with a higher $h$-index than M and almost as high an $h_\alpha$ index, is not a Distinguished Professor.
Comparing N with C and G, it seems incomprehensible that N, with $much$ higher $h$ $and$ $h_\alpha$ than C  and G, is at a lower academic rank than  C and G. 
In this author's opinion, that is informed by detailed knowledge of the scientific record of these physicists, 
these inconsistencies are not a reflection of shortcomings of the bibliometric indices
$h$ and $h_\alpha$ to quantify scientific achievement, but rather reflect the failure  of the
academic promotion process at this major research university to properly reward higher scientific achievement
with a higher academic rank for the scientist and vice-versa. 

Physicist AA in  table I has a high $h$-index, but not qualitatively different from that of  others on the list, however has a remarkably
high $h_\alpha$ as well as ratio $r_\alpha=h_\alpha/h=0.93$,   the highest in the group by a large margin. We have found  such high values of $r_\alpha$ 
 only among exceptionally
accomplished scientists that have earned broad acclaim. Physicist AA is a Nobel laureate.
 
\section{more examples}

In table II we list  the bibliometric data of 10  mid-career active condensed matter theorists \cite{table2}.
 Their ``age'' (i.e.  time since their first paper) ranges from 11 to 26 years, mostly clustered around 20 years, and their $h-$indices 
 range from 16 to 32.  No systematic rule was used in choosing these examples, other than keeping $h$ and `years' within 
 limited ranges, and choosing scientists where either themselves and/or their   coauthors were known to the author of this paper, to 
 facilitate the process of finding their $h_\alpha$-index.
 We have also computed $h'_\alpha$ for these scientists, which was considerably more time-consuming than computing $h_\alpha$.

 As expected there is not a strong correlation between ``age'' and $h-$index in table II, in other words    $m=h/years$ values
 vary widely, ranging from 0.75 to 1.82. 
 None  of these scientists works in large collaborations, their papers have typically one to a few coauthors. 
 The average number of coauthors for papers in their $h-$core ranges from 1.3 to 3.5 as shown on table II, 2.7 is the overall average.
 Nevertheless   their $h_\alpha$'s are very different,  hence so is their ratio $r_\alpha$. 
 
 Let us start with physicist T, with $h=20$ and the smallest $h_\alpha=r_\alpha=0$. This is the youngest of the group, with the highest $m=1.82$ and one of
 the largest total citation numbers. The impressive citation metrics (excluding $h_\alpha$)  come from
 collaborations with much more senior highly cited physicists such as D. Scalapino (h=97), S.C. Zhang (h=90), D. A. Huse (h=85), M.P.A. Fisher (h=73),
 S. Kivelson (h=62), F. Haldane (h=55), S. Kashru (h=50), 
 S. Chakravarty (h=49). It would be difficult to believe that T is the leader in   these collaborations.
 Even looking beyond the $h-$core, T has no single-author
 papers and only a handful of papers with few citations where T is the $\alpha-$author. So in comparing T's
 bibliometric record  with that of other
 physicists it would be misleading to $not$ consider $h_\alpha$. In the absence of $h_\alpha$ one would conclude from the bibliometric information that T is the most accomplished physicist 
 in table II. Instead, knowing that $h_\alpha=0$, at least indicates that one has to have a closer look.
  One may assume  that the reason $h_\alpha=0$ for physicist T is because T is very junior, and in the future 
 $h_\alpha$ and $r_\alpha$ will increase. This is suggested by the fact that T's $h'_\alpha=6$, mostly from recent papers. The future will tell.  
 
 At the other extreme we have physicist V, with comparable $h-$index to T, $h=25$ vs. $h=20$, a substantially smaller $m=1.25$
 but  remarkably high $h_\alpha=18$ and $r_\alpha=0.72$. V does have  a few papers with scientists with higher $h$ including very senior scientists (N. Ashcroft  (h=60), H. Kleinert (h=40), A. Sudbo (h=39)). However, V  has a considerable number of
 highly cited single author papers  (6 papers out of 25 in the $h-$core) and many highly cited papers with junior coauthors,
 resulting in the very high $h_\alpha$ and alpha-ratio $r_\alpha$. V's  $r_\alpha$ is also  the highest among all the entries in
 table I other than AA, despite being more junior than 13 out of the 17 physicists on that list. These data suggest
 remarkable independence and scientific leadership for this relatively young physicist.

 \begin{table}
\caption{Bibliometric data for 10 condensed matter theorists of comparable age and $h-$indices.
`coauth' is the average number of coauthors for papers in the $h-$core of the author.
$r_\alpha = h_\alpha/h$, $m'_\alpha = h'_\alpha/years$
}

\begin{tabular}{l || c | c | c | c| c|c| |c |c | c|c }
& $h$ & $h_\alpha$ & $r_\alpha$ & $h'_\alpha$ & $m'_\alpha$& m& pubs &citations &yrs& coauth  \cr
\tableline
\tableline
Q&16 &    \bf{1}  &\bf{0.06} & \bf{4}& \bf{0.20} &\bf{0.80}& 38&895& 20 & 3.3 \cr
\tableline
R&17 &    \bf{5} &  \bf{0.29} &\bf{9}&\bf{0.53} &\bf{1} & 35 &1032& 17 & 2.2 \cr
\tableline
S&17 &    \bf{8} & \bf{0.47} &\bf{10}&\bf{0.56} &\bf{0.94} & 51&1590& 18 & 3.5  \cr
\tableline
T&20 &    \bf{0} & \bf{0.00} &\bf{6}&\bf{0.55} &\bf{1.82} & 54&3468& 11 & 2.9  \cr
\tableline
U&22 &    \bf{3} & \bf{0.14} & \bf{4}&\bf{0.19} &\bf{1.05} &  40&3531& 21 & 2.7  \cr
\tableline
V &25 &    \bf{18} & \bf{0.72} &\bf{21}&\bf{1.05} & \bf{1.25} & 75&2096 & 20& 1.3 \cr
\tableline
W &27 &    \bf{6} & \bf{0.22} &\bf{12}&\bf{0.46} &  \bf{1.04} & 109 &2349  & 26& 2.5  \cr
\tableline
X &28 &    \bf{7} & \bf{0.25} &\bf{16}&\bf{0.70} &  \bf{1.22} & 90&2590 & 23& 3.5  \cr
\tableline
Y &31 &    \bf{17} & \bf{0.55} &\bf{21}&\bf{1.24} &  \bf{1.82}  & 95&2616 & 17& 2.7 \cr
\tableline
Z &32 &    \bf{20} & \bf{0.63} &\bf{26}&\bf{1.30} &  \bf{1.60}  & 114&3059 & 20& 2.1  \cr
\tableline

\end{tabular}
\end{table}

 It is also apparent from table II that $r_\alpha$ is not strongly correlated with age. Physicist Q, of the same age as V,
 has the second smallest $r_\alpha=0.06$, and the oldest physicist in the list, X, has a relatively small $r_\alpha=0.21$. 
 The physicists with highest $r_\alpha$ on this list are S, V, Y, Z,  with $r_\alpha=0.47, 0.72, 0.55$ and $0.63$ respectively
 and ages in the mid-range, $18, 20, 17, 20$ respectively. It is however the case that it is rare to find
 physicists with smaller $h$ that have a large $r_\alpha$, like physicist S on this list.  As $h-$indices become larger,
 larger values of $r_\alpha$ become increasingly more common. 

As expected, $h_\alpha$ and $h'_\alpha$ give similar information. Do we learn anything new from $h_\alpha'$? Yes we do.
Recall that $h'_\alpha$ also counts the $\alpha$-papers that are not in the $h$-core.
For example, comparing T and U, both have similar $h$ and $h_\alpha$. However, T has $h_\alpha=0$, $h'_\alpha=6$, while U has $h_\alpha=3$, $h'_\alpha=4$.
This indicates that T has several $\alpha-$papers not yet  in the $h$-core  with appreciable number of citations,
which are likely to enter the $h-$core in the near future and at that point increase   T's $h_\alpha$. In contrast, the fact that
$h'_\alpha - h_\alpha=1$ for U indicates that U does not have many papers with appreciable citations
that are not in the $h-$core. This suggests that T's $r_\alpha$ is likely to be larger than U's $r_\alpha$ in the near future. Thus, 
while comparing the bibliometric data of T and U  including $h_\alpha$ but not $h'_\alpha$ may suggest that U is more accomplished, taking $h'_\alpha$ into account
reverses this conclusion. A  large difference between $h_\alpha$ and $h'_\alpha$, as seen in table II for authors T and X,
indicates  that the author is becoming increasingly independent and increasingly leading   his/her research efforts.

We also list in table II the ratio $m'_\alpha=h'_\alpha/years$, which gives the same information as 
$m=h/years$ but only for the papers where the author is the $\alpha-$author.  $m$ and $m'_\alpha$  give  measures of the
scientist that are independent of his/her seniority. We argue that $m'_\alpha$ gives a truer measure of the scientist than $m$  because
it is less dependent on coauthorship patterns. For scientists where there is a large difference between
$m$ and $m'_\alpha$, such as Q, T, U, W, we suggest that this  raises concern about how much the value of $m$ is a true reflection 
of the scientist.

 We believe that the bibliometric information in table II, as well as in table I,  clearly illustrates
  the importance of taking into account the proposed index $h_\alpha$ and ratio $r_\alpha$, and if available,
  also $h'_\alpha$ and $m'_\alpha$, to complement the bibliometric information given by $h$ and $m$.
  Scientists Q and U look substantially weaker when the $\alpha-$information is taken into account than in its absence.
From the values of $m$ in table II one would conclude that physicists T, Y, Z, V  are the most accomplished
  on the list, in decreasing order. Instead, according to $r_\alpha$, it is V, Z, Y, S,
  and according to $m'_\alpha$ it is Z, Y, V, X. Thus, S, T and X do not excel according to all these criteria, while V, Y and Z do, 
  which is reassuring. The situation is rather different in table I, where several scientists that excel according to $m$
  do not excel at all according to $r_\alpha$, nor presumably according to $m'_\alpha$. 
  Of course, detailed examination of all these authors' publication records and other information could change these conclusions.

  Note that the number of coauthors of the scientists in table II are all similar. This indicates 
  that the fractional allocation schemes proposed in Refs. \cite{schr1,hh1,hh3} would reduce all the
  $h$-indices by similar factors in obtaining the $h_m$, $fractional$ $h$ and $gh$ indices.
  Thus, those indices would not reflect the different leadership patterns reflected in the
  $h_\alpha$ index. As an example, scientists W and Y have h-indices 27 and 31, and number of 
  coauthors 2.5 and 2.7. The ratio is 27/2.5=10.8, 31/2.7=11.5, very similar, yet their $h_\alpha$ of
  $6$ and $17$ is very different. Similarly the large differences in $h_\alpha$ and $r_\alpha$ of
  the scientists in table I is not accounted for by different number of coauthors. With the
  exception of   G, K, O and P, all the other scientists in table I have a small number of
  coauthors similar to those in table II.

 \section{technical details}
 
 Let us illustrate the procedure we use to obtain $h_\alpha$  in more detail for the case of physicist AA in table I, where it is particularly easy because a large fraction
of AA's paper are single author or with very few coauthors. This will also allow us to suggest capabilities that could be incorporated in the existing
bibliometrics databases to make the calculation of $h_\alpha$  simpler. We will use the Web of Science.

Physicist AA is F.D.M. Haldane, he has authored 116 papers, 35  of which are single author, a remarkably high number compared to typical condensed matter theorists. 
Even more remarkably, 25 single-author papers are in his h-core, and 8 of his 10 most highly cited papers are single author.
This alone demonstrates his remarkable independence and scientific leadership. In collaborative work he is almost always
the $\alpha-$author, resulting in his very high $r_\alpha=h_\alpha /h=0.93$ ratio. 

To find his $h_\alpha$ index we go through the list of his publications in order of decreasing citations. The first non-single-author
paper is paper 7, coauthored with S. Raghu, with 768 citations. Clicking on the paper title, then on Raghu's name, then
on "Create Citation Report" for Raghu, we learn that Raghu's $h-$index is 21, smaller than Haldane's  55, hence
this paper contributes to Haldane's $h_\alpha$ index. Continuing down the list of Haldane's publications we find the next 
collaborative paper is paper 9
with H. Li, also contributing to Haldane's $h_\alpha$ since Li's $h-$index is smaller than 55. The next collaborative paper is number
13, with I. Affleck, whose $h-$index is 73, larger than Haldane's 55, therefore this paper does not contribute to Haldane's
$h_\alpha$. Continuing this process we find many  papers with coauthors of $h-$index lower than Haldane's
(e.g. Rezayi, Arovas, Auerbach, Bernevig, Bhatt) that contribute to Haldane's $h_\alpha$, and papers with coauthors P.W. Anderson (h=108),  
P. Littlewood (h=61) and L. Balents (h=61) that have $\geq$ 55 citations (i.e. are in Haldane's h-core) and
do not contribute to Haldane's $h_\alpha$ because the  coauthors have $h$ larger than 55.
We continue this rather tedious procedure until reaching paper 56 in the publication list that has fewer than 55 citations,
at this point we stop and have found a total of 51 of the 55 papers in Haldane's $h$-core  that are $\alpha$-papers for Haldane, hence
his $h_\alpha$ index is 51 and his $r_\alpha$ is $51/55=0.93$.

 An alternative procedure would be  to click on the ``Analyze Results'' link in Haldane's  publication list, then click on ``Authors'' on the column
 on the left, to obtain the list of all of Haldane's coauthors, ordered by ``Record Count''. Next we would need to check
 the citation records of all the coauthors to find their $h-$indices. However, since we don't know from this page whether the coauthored papers are
 or are not in Haldane's $h-$core this is not an efficient procedure to find $h_\alpha$. If  the Web of Science were to provide the $h-$index of
 the coauthors on this page, and allow to order the coauthors in order of decreasing $h$, it would be very simple to
 find the coauthors with $h-$index larger than Haldane's, then look for the papers coauthored with these that are
 in Haldane's $h-$core, thus greatly simplifying the calculation of $h_\alpha$. 
 
 To obtain $h'_\alpha$ for Haldane, we continue down his publication list beyond paper 55. The next three are  
 $\alpha-$papers and have $\geq 53$ citations,  hence
 $h'_\alpha=53$, and $m'_\alpha=1.26$, incredibly close to $m=1.31$, which is a very unusual situation.
 Here, the extra work beyond computing $h_\alpha$ for  computing $h'_\alpha$ was negligible.
 However, for the cases in table II it took considerably longer to obtain $h'_\alpha$ because we had to 
 look at the citations of many papers not in the $h-$core of the scientist.

\section{summary and discussion} 

This paper was partially motivated by the increasingly wide use of the $h-$index to rank and compare scientists.
The shortcoming of the $h$-index to  differentiate scientists with different coauthorship  patterns already existed at the time
 the $h-$index was created, but we believe it may have been exacerbated
by the $h-$index itself in the ensuing years. There is no ``cost'' to the $h-$index of a scientist to work in large collaborations that include highly accomplished scientists, on the contrary,  there is potentially a large benefit in resulting in a higher $h$-index compared to the scientist pursuing his/her
own independent ideas in small collaborations or single-author papers. Thus this provides an incentive  for young scientists
to join large collaborations and/or collaborations with prestigious coauthors  even when there is not a compelling
scientific motivation for it, and we believe this may result in a non-optimal use of the scientists' abilities.

More generally, we have observed that there are many examples of scientists with comparable $h-$indices but very
different profile as far as scientific leadership is concerned, which we believe is a very important aspect of what is
generally understood as ``scientific accomplishment''.  Scientific advances result both from the contributions of scientific
{\it leaders} and scientific {\it followers}, but only the former are irreplaceable. We believe it is important to
identify and incentivize such scientists with proper citation metrics. The $h-$index alone increasingly doesn't do it
in this age of $h$-inflation.

To address these issues,  we aimed in this paper  to introduce a measure of the scientific production of a scientist that counts only those papers
where the scientist is the
leading author. Who that person is for a given paper
is a non-trivial question, except for 
single author papers. In some scientific subdisciplines, it is usually the last author in the author list. In others, it is
usually the first author.  Yet in others, authors are always listed alphabetically so the order of authors carries no
information. Is there a general criterion to identify this person? We proposed that the coauthor with the highest
$h-$index is the most likely candidate and called that author the $\alpha$-author, and the paper  an
$\alpha$-paper of that author.

One could argue that it would  be reasonable to use a more inclusive criterion to define what is an $\alpha-$paper of an author, that would allow
for more than one $\alpha-$author not only when top $h-$indices are identical.
For example, if the $h-$index of the author is within $10\%$ of the highest $h$-index of a coauthor, it may be argued that it is likely this
scientist also played a leading role, and count that paper as an $\alpha-$paper for that author also. Particularly for young scientists that collaborate
with peers of similar seniority without more senior coauthors we believe  that this would be a reasonable procedure.
One could call such an index $h_{\alpha xx}$, where $xx$ gives the percentage range for inclusion, i.e. $h_{\alpha 10}$ in the above example,
and similarly for $h'_{\alpha xx}$.
For the Haldane example, $h_\alpha=h_{\alpha 00}=51$, $h_{\alpha 10}=53$, $h_{\alpha 25}=54$, $h_{\alpha 50}=55=h$.

There will certainly be situations where our proposed criteria do not  reflect reality. For example, it is often the case that experimentalists that
make samples have very high $h-$indices. In an  experimental paper where such a sample is used  the scientist providing the sample would be the $\alpha$-author even if he/she  made otherwise no contribution to the scientific 
project, hence certainly didn't ``lead'' the project. Similarly, in a purely theoretical paper where an experimentalist supplied data hence is a coauthor,
a theorist with lower $h-$ index may well be the leading author in the project, yet not identified as such by our criteria.
These limitations underscore the fact that it is important to consider many factors besides bibliometric indices in the
evaluation of scientists.

We  defined the $h'_\alpha$ index in the same way as the $h-$index is defined, namely the number of $\alpha$-papers
written by a scientist that have $\geq$ $h'_\alpha$ citations. $h'_\alpha$ is the more consistent way to quantify the
criterion we are after. Unfortunately, $h'_\alpha$ is very time-consuming to obtain from the existing databases. 
One has to go through a large fraction of the papers of an author and find out whether or not it is an $\alpha-$ paper for
that author and whether or not it contributes to his/her $h'_\alpha$ index.  For senior scientists with many publications, citations, and coauthors, this is a very time-consuming process.
For that reason we defined the $h_\alpha$ index, which counts only
 the $\alpha$-papers of the author in the author's $h-$core, a subset of the papers in the $h'_\alpha$-core.
 The $h-$core is usually at least a factor of 3 smaller than the total number of papers of a scientist, thus
reducing the time necessary to compute the $h_\alpha$ index versus the $h'_\alpha$ index by a substantial factor.
$h_\alpha$ and $h'_\alpha$ have similar information.

A  drawback of $h_\alpha$ relative to $h'_\alpha$ is that there could be cases where the $h-$index of an author is  high principally because of multiauthored collaborations with scientists with even higher $h$-indices,  yet the
author may have quite a few  other papers not in his/her  $h$-core that contribute to a fairly high $h'_\alpha$ index. 
An example of this was scientist T in table II.
In such cases, which we believe are not very common, $h_\alpha$ could be much smaller than $h'_\alpha$ and give a somewhat distorted picture of the
author's scientific achievement and leadership. 

In fact, neither $h_\alpha$ nor $h'_\alpha$ are ideal definitions. Imagine that physicist O in table I, with $h=60$ and $h_\alpha=1$,
coauthors a paper with physicist B, with $h=27$ and $h_\alpha=19$. More likely than not, B would be the leader in the
collaboration, having shown leadership and/or independence 19 times before, versus 1 time for physicist O. However, according
to our definitions, O would be the $\alpha$-author having the higher $h-$index,  and the joint paper would 
potentially contribute to O's $h_\alpha$ but never  to
B's $h_\alpha$ (and the same for $h'_\alpha$). To avoid this, we could instead {\it define $\alpha-$ papers to be those papers
of an author  where the author has the highest $h_\alpha$ index} rather than $h$-index among the coauthors, 
with $h_\alpha$ as defined earlier, and 
a better index $h''_\alpha$ as:
{\it a scientist  has  index  $h''_\alpha$ if that scientist has written $h''_\alpha$ $\alpha$-papers with 
$\geq$   $h''_\alpha$ citations each}.  With these definitions, the paper coauthored by B and O would
be an $\alpha$-paper of B and not of O, hence  potentially contribute to the $h''_\alpha$-index of B but never  to that of O. 
Finally, for a self-consistent definition of $h''_\alpha$ we could define   a paper to be  an $\alpha-$paper of the
author that has the highest $h''_\alpha$ instead of the highest $h_\alpha$.  In any event,   the
non-self-consistent and even more so the  self-consistent $h''_\alpha$-indices would be very time consuming to obtain and 
for that reason of no practical
interest, certainly at present.

It is unquestionable that any given scientist, when looking at his/her own publication list, will find examples
where the $\alpha-$ author of a paper is not correctly adjudicated by our criterion, $and$ moreover where
all the other coauthors of the paper would agree. We argue that for any given scientist there will be papers
where he/she is not the $\alpha-$ author according to our criterion but should be, and also other papers
where the reverse is true, so that these errors would cancel out to zeroth order at least. For the overall 
publication record of a scientist we believe it would be very rare that $h_\alpha$ as defined here grossly
misrepresents the overall leadership role of the scientist, at least we have  not found any such examples
so far.

It has been noted that the criterion for $\alpha-$ author of a paper as defined here can change with time,
and argued  that this may be a  drawback for the $h_\alpha$ index \cite{priv}. We argue that this is not so. Consider
the situation where a junior  scientist, a student or postdoc, collaborates with his/her research advisor, a senior scientist. The paper will initially be
an $\alpha-$ paper of the senior scientist. If much later the junior scientist's h-index surpasses that of the
advisor, the paper will become an $\alpha-$ paper of the junior scientist and no longer
be an $\alpha-$paper of the advisor. That may simply reflect the fact
that this junior scientist is very talented and is likely to have played the leading role in this
early paper even if it was not initially reflected in its  $\alpha-$ status. So we argue that the
fact that the $\alpha-$ status of a paper can change with time is more likely to be an advantage rather than
a drawback.

Despite all  these caveats, we argue that the $h_\alpha$ index proposed in this paper and its variants provide an essential 
complement to the $h-$index. They can provide a clear distinction between scientists with similar $h-$indices but
very different coauthorship patterns, in particular distinguish  between 
scientists publishing with few coauthors and those in large collaborations, and distinguish between scientific leaders and followers. 
For two scientists with similar $h-$indices but very different $h_\alpha$ indices, we argue that it is highly likely
that the scientist with higher $h_\alpha$ index is the more accomplished one.  For two scientists with reverse ordering in $h$ and $h_\alpha$ the comparison
has to be done with care. If forced to choose between $h$ and $h_\alpha$ to rank scientists, this author believes
that in the absence of other information, $h_\alpha$ should carry more weight. However, both $h$ and $h_\alpha$ carry
important information and should be used together. If available, the index $h'_\alpha$ has additional important 
information that should also be considered.

The $h_\alpha$ index can be obtained with
moderate  work using the existing bibliometric databases,
and we argue that in assessing and comparing the achievements  of  scientists using bibliometric data one should $never$
do a comparison  using the $h$ index alone without also using the $h_\alpha$ index. Furthermore, for cases where $r_\alpha=h_\alpha/h$ is very small,
as in some of the examples seen, it is important to consider the additional information that is provided by
$h'_\alpha$ even if it involves additional substantial effort.

To the extent that consideration of $h_\alpha$ in the assessment of scientists  gains acceptance, we believe it will provide
additional incentive for young scientists to pursue  innovative work following their own
ideas, versus joining collaborations with more senior scientists and following their established  ideas that may not always be correct. We believe that this incentive would be beneficial to the vitality and innovative quality of the scientific enterprise.

In summary, we propose  that taking into account the $h_\alpha$ index    of a scientist
and $r_\alpha$ ratio in addition to his/her $h-$index and $m$-ratio, as well as $h'_\alpha$ and $m'_\alpha$ if available, should
result in better and fairer decisions regarding allocation of funding resources, career advancement of
scientists, decisions on scientific awards and elections to prestigious scientific bodies. To the extent that bibliometric 
databases such as Web of Science, Scopus and Google Scholar, introduce tools to facilitate the calculation of
$h_\alpha$ and $h'_\alpha$  indices, and even $h_{\alpha xx}$ and $h'_{\alpha xx}$ indices, as they have done for the $h-$index, we believe that this will have a positive effect on the
advancement of science.
\newline

\acknowledgements
The author is grateful to a colleague  for thoughtful comments.


\begin{references}
\bibitem{dist} There are six other theoretical physicists in this department at the rank of Distinguished Professor,
all have $h-$indices higher than 50.
\bibitem{table2} Scientists interested to know whether they are listed in table II can obtain this information from the author
upon request.
\end{references}
\end{document}